\documentclass[pdflatex,sn-mathphys-num]{sn-jnl}

\usepackage{graphicx}%
\usepackage{multirow}%
\usepackage{amsmath,amssymb,amsfonts}%
\usepackage{amsthm}%
\usepackage{mathrsfs}%
\usepackage[title]{appendix}%
\usepackage{xcolor}%
\usepackage{textcomp}%
\usepackage{manyfoot}%
\usepackage{booktabs}%
\usepackage{algorithm}%
\usepackage{algorithmicx}%
\usepackage{algpseudocode}%
\usepackage{listings}%
\usepackage{float}
\usepackage{pdfpages}

\theoremstyle{thmstyleone}%
\theoremstyle{thmstyletwo}%

\theoremstyle{thmstylethree}%

\begin{document}

\title[Article Title]{Oxide Interface-Based Polymorphic Electronic Devices for Neuromorphic Computing

}


\author*[1]{\fnm{Soumen} \sur{Pradhan}}\email{soumen.pradhan@uni-wuerzburg.de}

\author[1]{\fnm{Kirill} \sur{Miller}}

\author*[1]{\fnm{Fabian} \sur{Hartmann}}\email{fabian.hartmann@uni-wuerzburg.de}
\author[2]{\fnm{Merit} \sur{Spring}}
\author[2]{\fnm{Judith} \sur{Gabel}}
\author[2]{\fnm{Berengar} \sur{Leikert}}
\author[1]{\fnm{Silke} \sur{Kuhn}}
\author[1]{\fnm{Martin} \sur{Kamp}}
\author[3]{\fnm{Victor} \sur{Lopez-Richard}}
\author[2]{\fnm{Michael} \sur{Sing}}
\author[2]{\fnm{Ralph} \sur{Claessen}}
\author[1]{\fnm{Sven} \sur{H\"{o}fling}}
\affil*[1]{\orgname{Julius-Maximilians-Universität Würzburg, Physikalisches Institut and Würzburg-Dresden Cluster of Excellence ct.qmat}, \orgdiv{Lehrstuhl für Technische Physik}, \orgaddress{\street{Am Hubland}, \city{Würzburg}, \postcode{97074}, \state{Bavaria}, \country{Germany}}}

\affil[2]{\orgname{Julius-Maximilians-Universität Würzburg, Physikalisches Institut and Würzburg-Dresden Cluster of Excellence ct.qmat}, \orgdiv{Experimentelle Physik 4}, \orgaddress{\street{Am Hubland}, \city{Würzburg}, \postcode{97074}, \state{Bavaria}, \country{Germany}}}

\affil[3]{\orgname{Universidade Federal de São Carlos}, \orgdiv{Departamento de Física}, \orgaddress{\city{São Carlos}, \postcode{13565-905 }, \state{SP}, \country{Brazil}}}


\abstract{Aside from recent advances in artificial intelligence (AI) models, specialized AI hardware is crucial to address large volumes of unstructured and dynamic data. Hardware-based AI, built on conventional complementary metal-oxide-semiconductor (CMOS)-technology, faces several critical challenges including scaling limitation of devices \cite{shalf2015computing,bespalov2022possibilities}, separation of computation and memory units \cite{tang2019bridging} and most importantly, overall system energy efficiency \cite{ref1}. While numerous materials with emergent functionalities have been proposed to overcome these limitations, scalability,
reproducibility, and compatibility remain critical obstacles \cite{singh2024brain,zhang20222d}. Here, we demonstrate oxide-interface based polymorphic electronic devices with programmable transistor, memristor, and memcapacitor functionalities by manipulating the quasi-two-dimensional electron gas in LaAlO$_3$/SrTiO$_3$ heterostructures \cite{ohtomo2004high,thiel2006tunable} using lateral gates. A circuit utilizing two polymorphic functionalities of transistor and memcapacitor exhibits nonlinearity and short-term memory, enabling implementation in physical reservoir computing. An integrated circuit incorporating transistor and memristor functionalities is utilized for the transition from short- to long-term synaptic plasticity and for logic operations, along with in-situ logic output storage. The same circuit with advanced reconfigurable synaptic logic operations presents high-level multi-input decision-making tasks, such as patient-monitoring in healthcare applications. Our findings pave the way for oxide-based monolithic integrated circuits in a scalable, silicon compatible, energy efficient single platform, advancing both the polymorphic and neuromorphic computings.
}




\maketitle

In the era of advanced artificial intelligence (AI) technology, the increasing volumes of data generated and updated in our daily lives have created high demand for fast and energy efficient computing systems \cite{jones2018stop}. However, training of AI models requires enormous amounts of energy \cite{crawford2024world,sevilla2022compute}, especially when implemented in traditional complementary metal-oxide-semiconductor (CMOS) technology-based electronic hardware, due to limitations in device scaling \cite{shalf2015computing,bespalov2022possibilities}. As a promising alternative, polymorphic technology---a special class of reconfigurable technologies---is capable of reconfiguring its hardware functionality irrespective of time and space \cite{hentrich2011polymorphic}, simplifying circuitry by reducing the number of electronic components, thereby decreasing both the area and energy consumptions \cite{raitza2017exploiting}. On the other hand, to overcome the von Neumann bottleneck \cite{tang2019bridging}, human brain-inspired neuromorphic computing has gained significant attention as a new computing paradigm that offers parallel signal processing with low energy consumption \cite{tang2019bridging,yang2019memristive}. Reservoir computing (RC), as one of them, needs training only on the `readout function' to produce a desired output which significantly reduces the training costs and can be employed in state-of-the-art hardware prototypes for pattern recognition \cite{danial2019two}, signal processing in noisy environments \cite{du2017reservoir} and unsupervised learning \cite{shchanikov2021designing}. Notably, neuromorphic transistors have the ability to reconfigure logic operations \cite{wang2023boolean} and store the logic output \cite{stone1970logic,wang2023boolean}, enabling their decision making capabilities \cite{choi2023physically} and applications in adaptive learning \cite{fuller2019parallel}, and edge computing \cite{chen2019cmos}. 

Till today, extensive research has been conducted on neuromorphic computing based on a wide range of material systems, including two-dimensional (2D) materials \cite{cao20212d,liu2020two}, inorganic compounds \cite{huang2020zero}, organic materials \cite{dai2018light} and oxides \cite{chen2019solar}. Among them, 2D-materials-based neuromorphic devices have attracted significant attention due to their exceptional electronic and optical properties \cite{goswami2023fabrication,singh2022pulsed,tang2022tunable}. Additionally, 2D materials have emerged as promising candidates for reconfigurable technologies \cite{peng2023programmable,tsai2023reconfigurable}, alongside silicon nanowires \cite{weber2014reconfigurable}. However, structural complexity of most of the reported devices raises the manufacturing costs. Also, these devices face challenges, including wafer-scale fabrication, integration with existing technologies, performance variability, and most importantly, degradation in air, which impacts long term stability \cite{singh2024brain,zhang20222d}. In contrast, oxide materials continue to garner interest as they effectively mitigate many of these limitations, offering enhanced durability, scalability and compatibility with conventional fabrication processes \cite{yoo2024efficient,rao2023thousands,demasius2021energy}.

In this context, the discovery of a highly mobile quasi-two-dimensional electron gas (q2-DEG) at the interface of LaAlO$_3$/SrTiO$_3$ (LAO/STO) heterostructures has paved the way for development of oxide-based next-generation electronic devices \cite{ohtomo2004high,mannhart2008two}. However, the manipulation of the q2-DEG, using metallic top-gate electrodes, results in additional band bending at the interface \cite{bi2016electro,goswami2015nanoscale,giampietri2017band}, while back-gate electrodes are limited to global control of the q2-DEG and often require cryogenic temperatures \cite{choe2019gate}. Hence, a viable and better alternative is laterally defined side-gates alongside a nanowire channel, which can be processed easily on the surface and have shown considerable potential in LAO/STO heterostructures \cite{schneider2006microlithography,miller2021room,monteiro2017side,stornaiuolo2014weak}. Beyond conventional field effect functionalities, employing side-gates also leads to the emergence of memristive operation, attributed to the migration of oxygen vacancies or floating gate effects \cite{miller2021room,maier2017gate}. Interestingly, the memristive properties of these devices can be intrinsically linked to memcapacitive behavior \cite{5247127}. However, only a few studies have reported hysteresis in capacitance in LAO/STO heterostructures ascribed to oxygen vacancies migration \cite{wu2013electrically}, structural distortion \cite{bi2016electro} or interfacial trap state \cite{kim2015electric}. 

In this study, we present a polymorphic electronic device that leverages the q2-DEG at the LAO/STO interface, capable of operating as a transistor (T), memristor (M), or memcapacitor (MC) at room temperature, depending on the biasing condition. By integrating 1T with 1MC, a RC system is implemented using a 4bit pulse scheme. Furthermore, combining 2T and 1M, logic OR and AND operations are demonstrated, with the significant advantage of in-situ data storage within the computing circuit. Notably, both logic functions are reconfigurable within a single circuit design, allowing its implementation in programmable operating systems. These versatile functionalities along with the broad range of combinatorial applications highlight the device potential for further advancing the field of oxide electronics.

\section*{Polymorphic electronic devices}
\begin{figure}[H]
\centering
\includegraphics[width=0.9\textwidth]{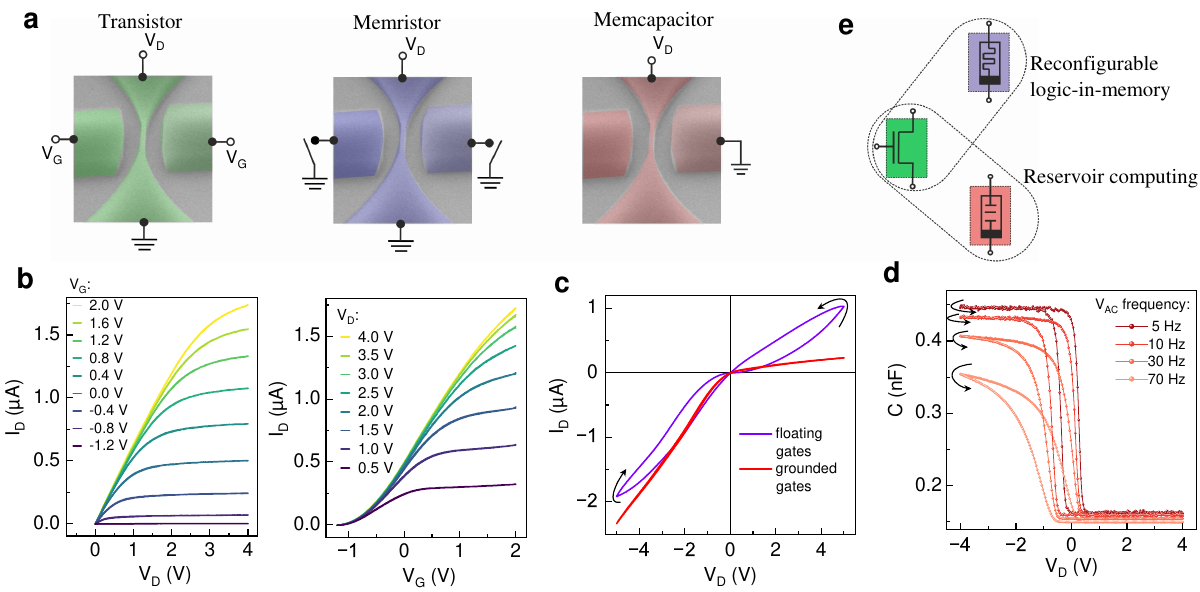}
\caption{\textbf{Polymorphic functionality of a runtime reconfigurable oxide interface-based device:} \textbf{a}, false-colored electron microscope images of the device alongside the circuit diagram for transistor (left panel), memristor (middle panel), and memcapacitor (right panel) operation modes. \textbf{b}, Output characteristics of the device for different gate voltages, $V_{\text{G}}$ (left panel), and transfer characteristics for different drain voltages, $V_{\text{D}}$ (right panel). The device operates as a planar n-channel field-effect transistor. Dependent on the gate potential, the channel can be depleted. \textbf{c}, $I_{\text{D}}$ ($V_{\text{D}}$)-characteristics for floating and grounded gates. The device can be operated as a memristor (type-I) or a non-linear resistor depending on the gate potential. \textbf{d}, Capacitance-voltage-characteristics measured between the wire and one of the gates, while the other gate is left floating. The arrows indicate the direction of the sweep cycles. \textbf{e}, Schematic illustration showing reservoir computing applications integrating transistor-memcapacitor devices and reconfigurable logic-in-memory applications in a transistors-memristor integrated circuit.}
\label{fig1}
\end{figure}

We first focus on the polymorphism of a single device based on different wiring configurations. Figure \hyperref[fig1]{1a} depicts the false-colored scanning electron microscope images of the device alongside the corresponding circuit diagrams for field effect transistor (FET), memristive, and memcapacitive functionalities on the left, middle, and right panels, respectively. The left panel of Fig.\hyperref[fig1]{1b} shows the output characteristics for varying gate voltage ($V_{\text{G}}$). As drain voltage ($V_{\text{D}}$) increases, the drain current ($I_{\text{D}}$) exhibits saturation behavior similar to an n-channel FET. The transfer characteristics (right panel of Fig.\hyperref[fig1]{1b}) quantify the device's response to varying gate voltage at different $V_{\text{D}}$ values. The results demonstrate an on-current in the $\mu$A range and the channel can be fully depleted below -1 V. 

To switch the device functionality to memristor, the full cycle $I_{\text{D}}$ ($V_{\text{D}}$)-characteristics are examined for grounded and floating gates configurations, showing their dependency on gate potential (Fig.\hyperref[fig1]{1c}). When the side-gates are grounded, the $I_{\text{D}}$ ($V_{\text{D}}$)-curve exhibits no memory effect and no hysteresis. In contrast, with the gates floating, a pinched hysteresis loop emerges with two distinct resistance states near zero bias voltage, indicative of memristive response. This difference arises because grounding the lateral-gates allows any charge transferred to the gates to be drained, preventing charge accumulation and, subsequently, the development of hysteresis. Conversely, in the floating gate configuration, the side gates accumulate or release charges gradually, depending on drain voltage, which ensures the hysteresis in the full cycle map for an adequate voltage sweeping rate \cite{miller2021room}. A very high resistance ratio ($R_{\text{high}}$/$R_{\text{low}}$) of $\sim$5567 is observed around zero bias, corresponding to a type-I memristive response, that according to Ref.\cite{silva2022ubiquitous} is expected to arise under polarity-dependent charge accumulation or release. When a symmetric voltage sweep is applied with respect to the middle of the nanowire, a type-II memristive response can be observed (Supplementary Note 1 and Fig.S1). 

The third functionality of the device emerges from its memcapacitive response with hysteresis in capacitance-voltage-characteristics along with a distinct transition between high and low capacitance states (Fig.\hyperref[fig1]{1d}). The capacitive response observed here differs from previous reports that attributes to structural distortion \cite{bi2016electro}, large geometric capacitance \cite{li2011very}, or oxygen vacancy migration \cite{wu2013electrically}. In this device, the layout and the dynamics of charge trapping and detrapping provide the explanation for the capacitive transitions. Here, the device operates similar to a metal-oxide-semiconductor diode, where the capacitance exhibits an asymmetric response (Supplementary Fig.S2). Under accumulation conditions (reverse bias), the capacitance saturates at its higher value, corresponding to $C_{\text{high}}$$\sim$1/$d_\text{B}$, where $d_\text{B}$ is the distance between the channel and the gate. Under depletion conditions, the capacitance decreases to $C_{\text{low}}$$\sim$1/($d_\text{B}$+$d_\text{D}$), where $d_\text{D}$ represents the depletion length. Building on these primary functionalities, new integrated operations such as RC and reconfigurable logic-in-memory architectures are explored as schematically represented in Fig.\hyperref[fig1]{1e}, which will be discussed in the following sections.

\section*{Reservoir computing}
\begin{figure}[H]
\centering
\includegraphics[width=1\textwidth]{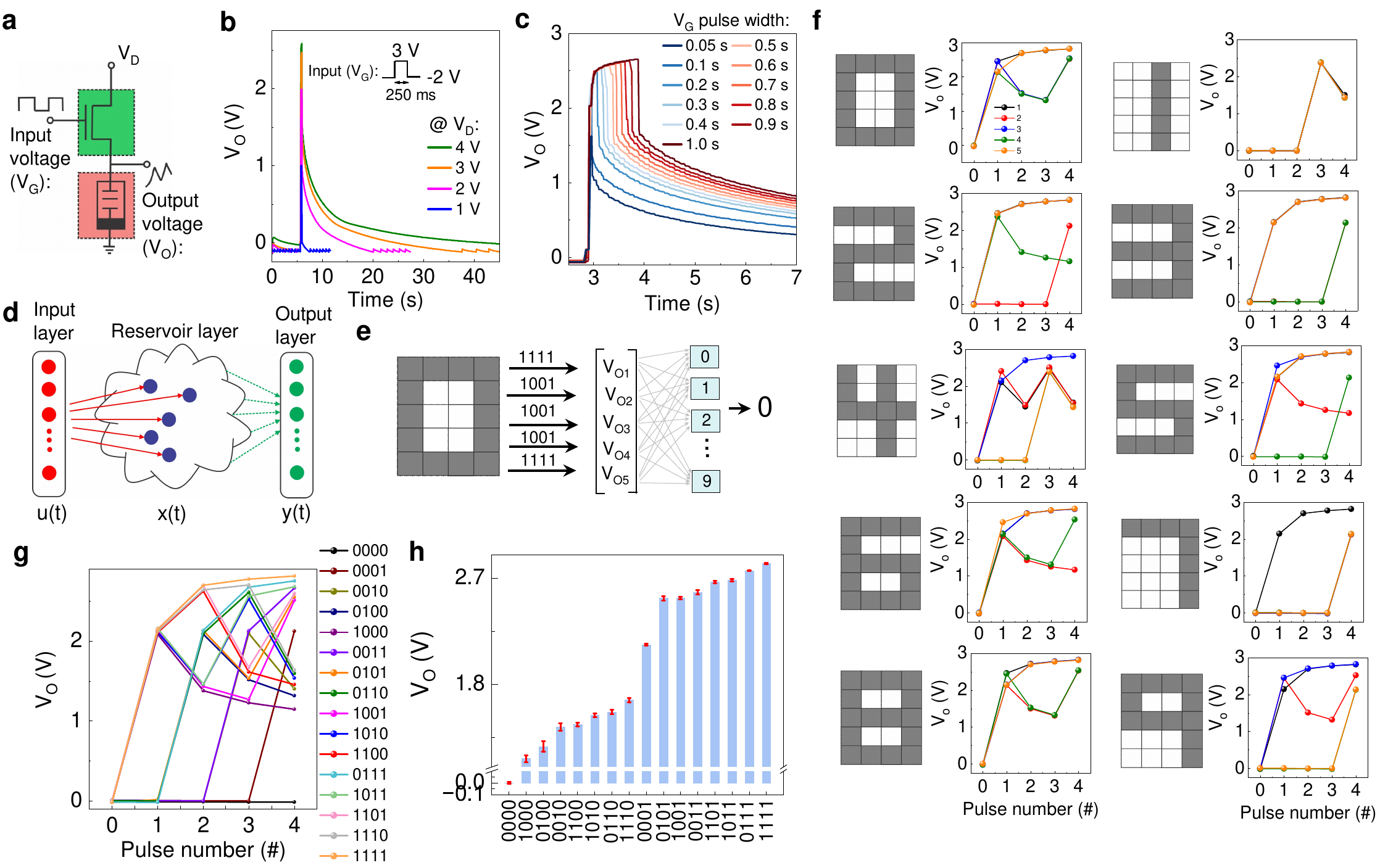}
\caption{\textbf{Reservoir computing operation in an integrated circuit with one transistor and one memcapacitor:} \textbf{a,} Schematic illustration of a 1T1MC integrated circuit with input voltage pulse to the gate ($V_{\text{G}}$) of T and output voltage ($V_{\text{O}}$) from MC while a drain voltage ($V_{\text{D}}$) is applied to the drain terminal of T; synaptic $V_{\text{O}}$ when applying a single input pulse between -2 and 3 V with \textbf{b,} a fixed pulse width of 250 ms and varying $V_{\text{D}}$, and \textbf{c,} varying pulse widths from 0.05 s to 1 s and fixed $V_{\text{D}}$ of 4 V; \textbf{d,} schematic of a reservoir computing  (RC) system with the reservoir layer directly connected to the output layer and only the output layer being trained to build the weight matrix, \textbf{e,} experimental demonstration of pattern recognition of a 0 digit image of 5$\times$4 pixels with an array of 5-reservoir outputs for pattern classification, \textbf{f,} $V_{\text{O}}$ for each pulse stream corresponding to each row in the digit images from 0 to 9, \textbf{g,} $V_{\text{O}}$ for all 16-types of 4-bit pulse trains and \textbf{h,} average $V_{\text{O}}$ with error bars at the end of all 16 states considering cycle to cycle variation.}\label{fig2}
\end{figure}

First, we demonstrate a neuromorphic computing architecture for RC application by connecting one transistor and one memcapacitor (Fig.\hyperref[fig2]{2a}). Notably, compared to transistor- and memristor-based RC systems where the reservoir state is defined by output current, memcapacitor-based devices offer key advantages: the voltage output eliminates static current induced power dissipation and the need for current-to-voltage conversion \cite{pei2023power}. To investigate the RC operation, an input voltage pulse ($V_{\text{G}}$=3 V) is applied to turn `on' the transistor for different $V_{\text{D}}$ values. As depicted in Fig.\hyperref[fig2]{2b}, for $V_{\text{D}}$ of 1 and 2 V, the output voltage ($V_{\text{O}}$) reaches $V_{\text{D}}$, while in the case of 3 and 4 V it reaches 2.47 V and 2.57 V, respectively, which occurs due to the polarity inversion between the drain and gate (explained in Supplementary Note 2). Furthermore, $V_{\text{O}}$ does not remain constant over time-even after the transistor is turned `off'-instead $V_{\text{O}}$ decays gradually, indicating charge leakage from the memcapacitor. The decay confirms the presence of short-term memory in the device, with the memory window extending as $V_{\text{D}}$ increases. Now, keeping $V_{\text{D}}$ fixed at 4 V, we vary the input pulse width and show a representative time window in Fig.\hyperref[fig2]{2c} (full range in Supplementary Fig.S4). The results reveal a nonlinear increase of $V_{\text{O}}$ with prolonged memory retention as we increase the pulse width. Therefore, our device configuration is well-suited for physical RC applications, serving as an effective reservoir layer that fulfills both the non-linearity and short-term memory requirements. 

A physical RC system comprises of three layers, where the input layer is mapped to high-dimensional space through the reservoir layer, which directly feeds into the output layer for direct classification (Fig.\hyperref[fig2]{2d}). To demonstrate this concept, consider a monochrome image of n$\times$m pixels, where each row of pixels is input sequentially to an array of n 1T1MC devices, which acts as the reservoir layer. The output voltages from each device are fed into the classification network in the output layer, consisting of \textit{d}-output neurons, where \textit{d} represents the number of classification labels. For classification, the dot products of the reservoir output, represented as an n$\times$1 vector, is computed with the n$\times$d weight matrix. The neuron label corresponding to the maximum dot product is identified as the predicted final output. The output layer requires supervised training to optimize the weight matrix for accurate classification. To illustrate this approach, we perform a digit recognition task using computer-generated 5$\times$4 digit pixel images from 0 to 9. Figure \hyperref[fig2]{2e} illustrates the schematic for digit 0 as an example. Figure \hyperref[fig2]{2f} displays the digit images alongside the output voltages corresponding to each 4-pixel pulse stream. As observed, $V_{\text{O}}$ is progressing at each pulse ``1'', while decaying towards its initial state with each ``0'' pulse, resulting in distinct $V_{\text{O}}$ values at the end of each pulse train based on the input sequence. The collective reservoir state forms a unique 5$\times$1 matrix for each digit. The reservoir state can then be effectively classified in the output layer through training. Note that only 6, out of 16 possible 4-bit combinations are used to represent the monochrome image of 10 digits. To further explore the device response, all 16 possible 4-bit pulse trains were applied to the device from its resting state (Fig.\hyperref[fig2]{2g}). To assess the consistency of the device response, each 4-bit pulse train was cycled 10 times. The average $V_{\text{O}}$ after stimulation, along with error bars, indicates that almost  all configurations can be reliably distinguished (Fig.\hyperref[fig2]{2h}). Therefore, our devices can be utilized in hardware-based RC systems as reservoir layer for pattern recognition.

\section*{Synaptic plasticity}
\begin{figure}[H]
\centering
\includegraphics[width=0.9\textwidth]{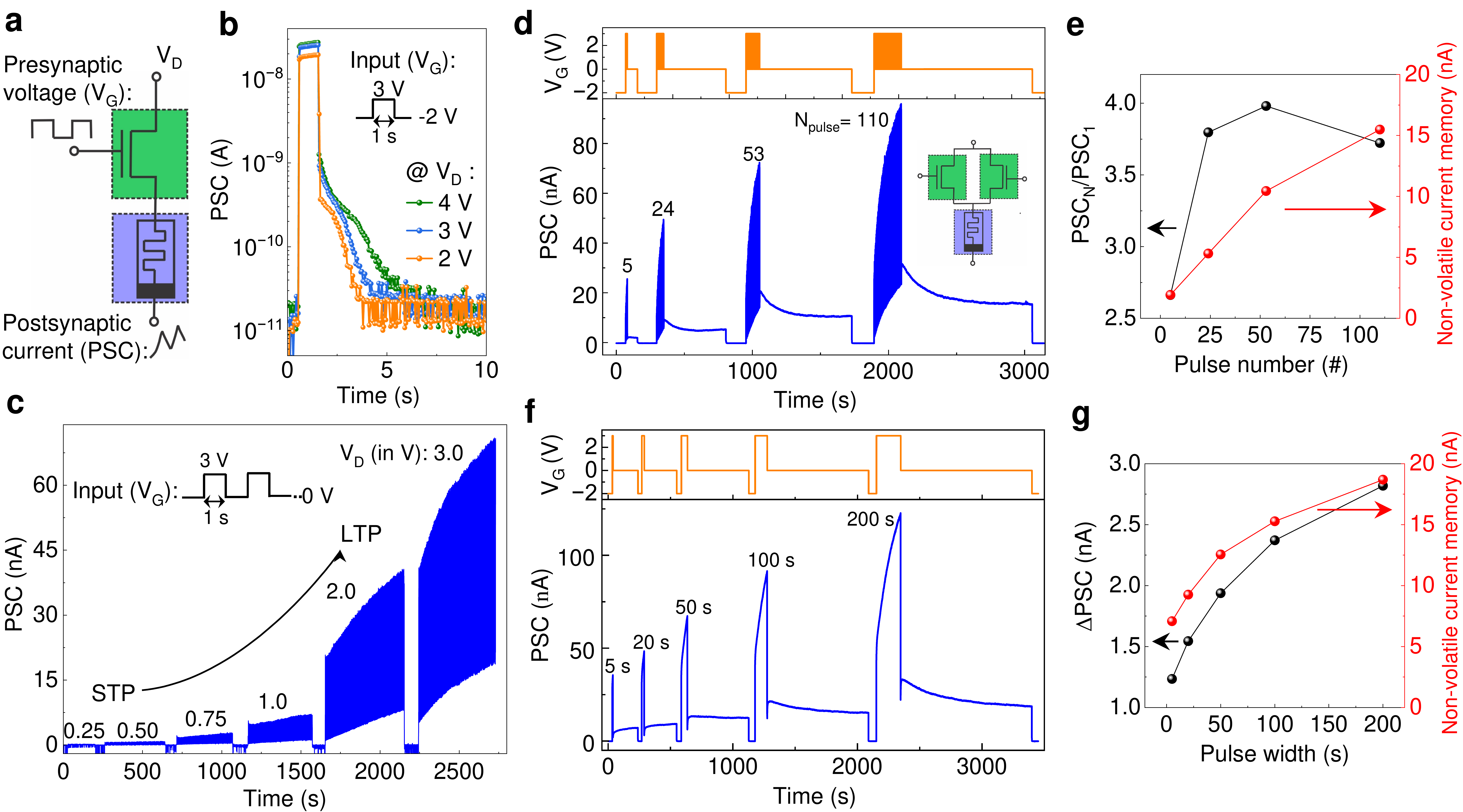}
\caption{\textbf{Short and long term synaptic operation by integrating transistor and memristor devices:} \textbf{a,} schematic diagram of a 1T1M combined circuit with input voltage pulse to the gate ($V_{\text{G}}$) of T and postsynaptic current (PSC) from M while the voltage ($V_{\text{D}}$) applied to the drain of T is read, \textbf{b,} short term potentiation (STP) in the circuit by switching `on' the transistor with a single $V_{\text{G}}$ pulse between -2 and 3 V at different $V_{\text{D}}$, \textbf{c,} post synaptic current (PSC) for continuous input voltage pulse train of 3 V with different $V_{\text{D}}$ from 0.25 V to 3 V, indicating the transition from STP to long term potentiation (LTP) with increasing $V_{\text{D}}$. Transition from STP to LTP in a 2T1M integrated circuit as seen in the PSC by changing \textbf{d,} the $V_{\text{G}}$ pulse number, \textbf{f,} the $V_{\text{G}}$ pulse width for $V_{\text{D}}$= 4 V. Inset in  \textbf{d} shows the schematic diagram of a 2T1M combined circuit with input $V_{\text{G}}$ pulse to the gates of both T and PSC from M. Spike number and width-dependent plasticity by showing \textbf{e,} variation of PSC$_N$/PSC$_1$ and non-volatile memory with pulse number,  \textbf{g,} variation of $\Delta$PSC and non-volatile memory with pulse width indicating higher non-volatile memory for longer pulse width and higher pulse number.}
\label{fig3}
\end{figure}

Artificial neural networks based on memristors are typically arranged in a crossbar layout. To address individual synaptic nodes and minimize sneak current paths into neighboring cells, a 1T1M configuration is employed, where the transistor acts as a selector \cite{bayat2018implementation} (Fig.\hyperref[fig3]{3a}), leveraging the device's inherent reconfigurability without compatibility issue. To investigate memory characteristics, a single $V_{\text{G}}$ pulse is applied to switch the transistor `on' for different $V_{\text{D}}$ values (Fig.\hyperref[fig3]{3b}). Here, the post synaptic current (PSC) rapidly reaches its maximum value, then retains an intermediate value for few seconds after the transistor is switched `off' before gradually returning to the baseline `off' current. This transient retention indicates short term potentiation (STP) within the device. Furthermore, Fig.\hyperref[fig3]{3c} displays PSC as a function of time under a train of $V_{\text{G}}$ pulses of amplitude 3 V for various $V_{\text{D}}$ values. At lower $V_{\text{D}}$, PSC remains unchanged despite successive pulses. However, at higher $V_{\text{D}}$, PSC progressively increases with each subsequent pulse demonstrating a transition from STP to long term potentiation (LTP). Additionally, the system can be reset to its initial state by fully turning `off' the transistor before proceeding to the next voltage condition. 

In the biological brain, continuous or repetitive stimulation of signals strengthens the synaptic weights, enabling the storage of information over extended periods, facilitating transition from STP to LTP. To mimic this operation in our device, synaptic properties are extracted by varying input pulse width and count for the 1T1M device configuration (Supplementary Note 4 and Fig.S6). For 2T1M circuit configuration, as schematically illustrated in the inset of Fig.\hyperref[fig3]{3d}, we performed similar experiments by applying identical $V_{\text{G}}$ pulses to the gates of both transistors. Figures \hyperref[fig3]{3d,f} depict the PSC as a function of time together with varying input pulse number and width, respectively, at $V_{\text{D}}$=4 V. The PSC increases non-linearly with pulse count and width, which can be attributed to the cumulative discharge from the floating gates to the nanowire. After each pulse scheme, the PSC initially decays exponentially before gradually saturating over time, indicating non-volatile memory characteristics. For better visualization of these results, we extracted the non-volatile current memory, spike-number-dependent-plasticity (PSC$_N$/PSC$_1$) and spike-width-dependent-plasticity ($\Delta$PSC) (Fig.\hyperref[fig3]{3e,g}). Initially, PSC$_N$/PSC$_1$ increases non-linearly with the number of pulses, but beyond 110 pulses, the ratio begins to decrease. This suggests that after a certain threshold, the charging rate on the floating gate increases compared to the discharging rate. However, $\Delta$PSC and the overall non-volatile memory increases with the number and width of pulses, confirming the transition from STP to LTP. This behavior demonstrates that our devices can efficiently emulate key functions of neuromorphic technology, enabling learning, temporary forgetting, relearning and long term information retention. 

\section*{Logic computation and its memory}
\begin{figure}[H]
\centering
\includegraphics[width=0.9\textwidth]{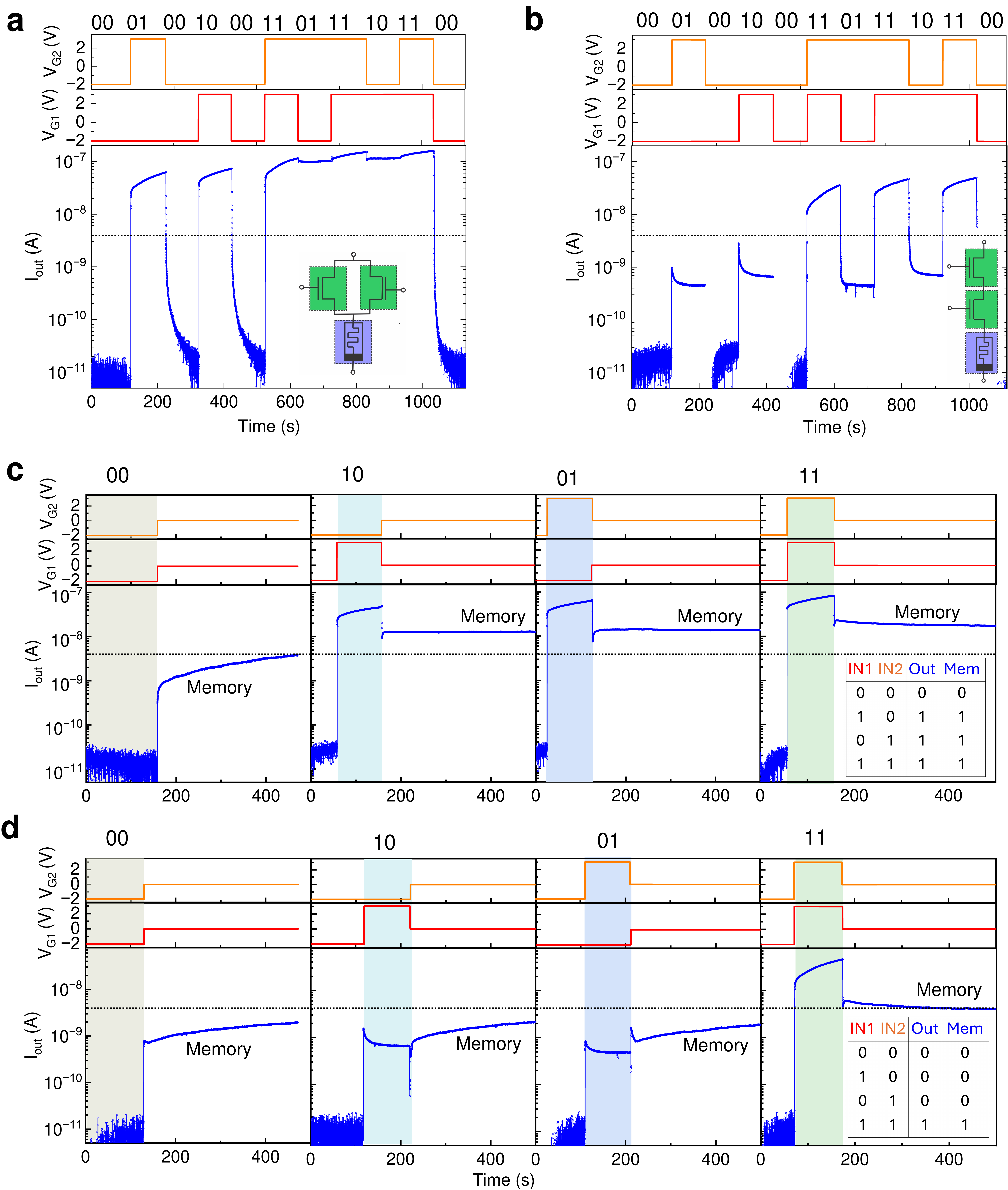}
\caption{\textbf{Logic OR and AND operations and their in-situ logic memory in two transistors/ one memristor integrated circuits: } different sequence of input signals and corresponding output signals demonstrating logic \textbf{a,} OR and \textbf{b,} AND operation and insets showing corresponding schematic diagrams of the 2T1M circuit configurations where the transistor gate voltages ($V_{\text{G}}$) are inputs and current is output signal ($I_{\text{out}}$); logic output signals during application of input signals and afterwards when the inputs are set to 0 V for \textbf{c,} OR logic operation in the circuit diagram shown in the inset of (a) and \textbf{d,} AND logic operation in the circuit diagram shown in the inset of (b). Insets on the right panels of (c) and (d) show the truth tables of logic output during procession of the inputs and memory afterwards for all input combinations for OR and AND logic gates, respectively. Here, $V_{\text{G}}$ of -2 V and 3 V are considered as logic input ``0'' and ``1'', respectively. The drain voltage ($V_{\text{D}}$) was kept fixed at 4 V during all logic inputs as well as for logic memory. The dashed lines show the threshold current of 4 nA to distinguish between logic output ``0'' and ``1''.}\label{fig4}
\end{figure}

In logic computation, OR, AND and NOT gates are fundamental building blocks, as all other logic gates can be constructed from them. NOT gate operation is demonstrated utilizing one T and one resistor (Supplementary Note 5 and Fig.S7). Here, we demonstrate OR and AND logic operations using 2T1M configuration. To perform logic operations, the transistors' gate voltages (denoted as $V_{\text{G1}}$ and $V_{\text{G2}}$) serve as input signals, while the current ($I_{\text{out}}$) through the memristor represents the output signal. Specifically, a negative voltage (-2 V) is defined as logic input ``0'', and a positive voltage (3 V) as logic input ``1''. The $I_{\text{out}}$ of 4 nA is considered as threshold between the logic output ``0'' and ``1''. For the OR gate operation, corresponding to the circuit shown in the inset of Fig.\ref{fig4}a, when both the inputs are logic ``0'', the $I_{\text{out}}$ is in pA range, representing output ``0'' (Fig.\hyperref[fig4]{4a}). For all other input combinations a significantly higher $I_{\text{out}}$ than 4 nA is observed, corresponding to logic output ``1''. This results in a clear distinction by 3 orders of magnitude between the ``low'' and ``high'' logic states, confirming the successful OR gate operation. Similarly, AND gate functionality is validated using the circuit shown in the inset of Fig.\hyperref[fig4]{4b}. Consistent with the OR gate operation, $I_{\text{out}}$ of few tens of pA and nA are observed  for ``00'' and ``11'' inputs, respectively (Fig.\hyperref[fig4]{4b}). For logic inputs ``10'' and ``01'', $I_{\text{out}}$ increases by one order of magnitude compared to “00”, but still remain below the threshold value. Therefore, the logic operation illustrated in Fig.\hyperref[fig4]{4b} demonstrates the functionality of an AND gate. To verify the consistency and reliability of the logic performance, all possible sequences of logic inputs are applied to both the OR and AND gate configurations. 

As previously discussed in Fig.\ref{fig3}, the 2T1M circuit configuration can also support long term memory through prolonged input stimulation. Here, the logic output memory is examined by monitoring the $I_{\text{out}}$ after setting the inputs to 0 V, following their stimulation for both the OR (Fig.\hyperref[fig4]{4c}) and AND (Fig.\hyperref[fig4]{4d}) logic. In all cases, immediately after input removal, there is a sudden fluctuation in $I_{\text{out}}$, but it does not cross the threshold value up to the memory examined time of 300 s. These results confirm that all logic outputs can be stored in-situ for both the logic gate operations. The underlying mechanism of logic memory retention is detailed in Supplementary Note 5 and Fig.S8. The insets in the right panels of Fig.\hyperref[fig4]{4c,d} present the corresponding truth tables for both the OR and AND gates summarizing the logic outputs during active input processing and their memory states. Therefore, the robustness of logic and in-situ memory of OR and AND gates in addition to NOT gate operation in the system display its potential in universal set of logic gate operations.

\section*{Reconfigurable synaptic logic}
\begin{figure}[H]
\centering
\includegraphics[width=1\textwidth]{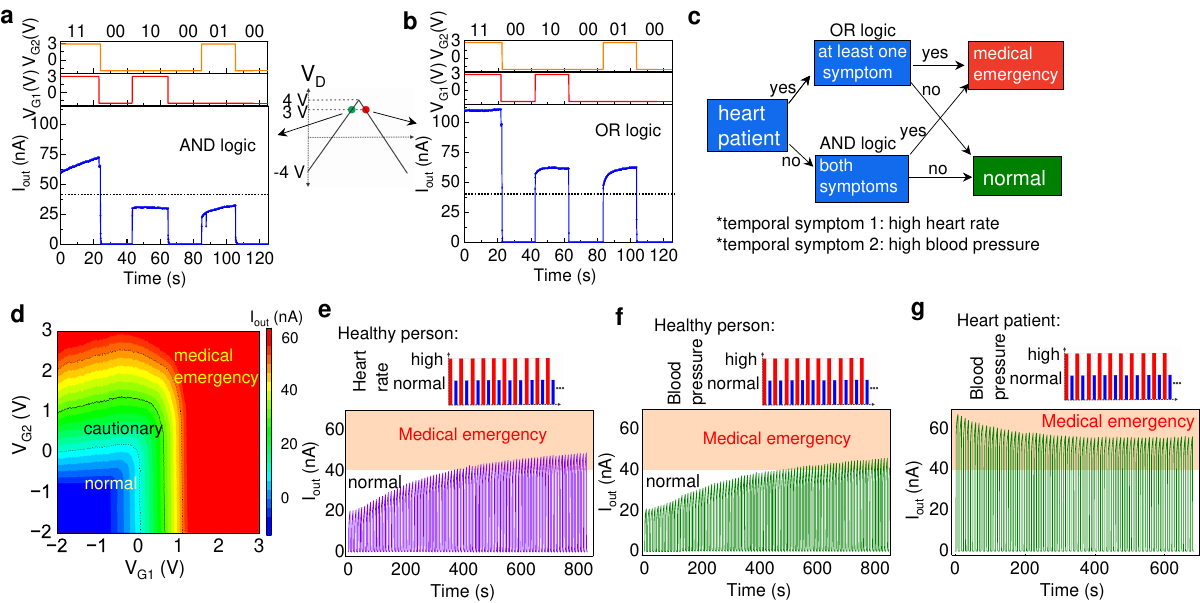}
\caption{\textbf{Reconfigurability of logic operations and its illustration for a healthcare task in a single circuit of two transistors and one memristor:} reconfigurable logic operation between \textbf{a,} AND and \textbf{b,} OR logic at $V_{\text{D}}$= 3 V after a $V_{\text{D}}$ sweep from -4 to 3 V and 4 to 3 V, respectively, for a 2T1M circuit configuration shown in the inset of Fig.\hyperref[fig4]{4a}; \textbf{c,} a diagnosing algorithm to identify a person's health condition considering a heart patient as static input and a high heart rate and a high blood pressures as non-static or temporal inputs, \textbf{d,} 2-dimensional current ($I_{\text{out}}$) map in the full range of input voltages ($V_{\text{G}}$) from -2 to 3 V assuming a normal-to-high range of heart rate (x-axis) and blood pressure (y-axis) at $V_{\text{D}}$=3 V indicating the device can be utilized as an analog type operator with three output cases (normal, cautionary and emergency cases); transition from normal to medical emergency case for a healthy person (static input) \textbf{e,} if blood pressure remains normal but heart rate continuously fluctuates and \textbf{f,} if heart rate remains normal but blood pressure continuously fluctuates; \textbf{g,} remaining at medical emergency condition for a heart patient (static input) if heart rate remains normal but blood pressure continuously fluctuates.}\label{fig5}
\end{figure}

 To mimic the decision making complex functionalities of the biological brain, one requires the ability to dynamically reconfigure logic computations based on external conditions. Here, the reconfigurability of logic operations is examined via the phase of the $V_{\text{D}}$ sweep direction for the 2T1M circuit shown in Fig.\hyperref[fig4]{4a} (Supplementary Note 7 and Fig.S9). We recorded $I_{\text{out}}$ over time at $V_{\text{D}}$=3 V for all input combinations following $V_{\text{D}}$ sweeps from -4 to 3 and 4 to 3 V for AND and OR logic operations, respectively (Fig.\hyperref[fig5]{5a,b}). It confirms that the phase of the $V_{\text{D}}$ sweep cycle effectively adds a new control dimension for logic computation and opens new possibilities for decision-making applications. For instance, these devices can be integrated into bioelectronic platforms to support diagnostic tasks \cite{choi2023physically}. Here, we design a simple diagnostic model to monitor two types of person with preconditions, healthy or heart-disease, and two diagnostic parameters: heart-rate and/or blood-pressure. The precondition is set in the phase of $V_{\text{D}}$, and thus either logic AND or OR applies. Real time diagnostic parameters are mapped to logic value inputs ``0'' and ``1'' (normal and elevated ranges of blood-pressure/heart-rate, respectively). As illustrated schematically in Fig.\hyperref[fig5]{5c}, for healthy individuals (AND logic), both diagnostic parameters need to be ``high'' to trigger a medical emergency warning. In contrast, for a heart-disease patient (OR logic), a medical emergency is triggered by the emergence of at least one elevated dynamic input. In all other cases the person is considered to be in normal state. Using this diagnosis algorithm, a person’s health status can be continuously monitored by comparing $I_{\text{out}}$ against the predefined threshold current. 

For the realization of an analog-type operation of the circuit, $I_{\text{out}}$ is mapped continuously by sweeping $V_{\text{G1,2}}$ between the initially used logic ``low'' and ``high'' voltage values (Fig.\hyperref[fig5]{5d}). Here, the map can be divided in three main regions, labeled ``normal'', ``cautionary'', and ``medical emergency'', depending on the two diagnostic parameters $V_{\text{G1,2}}$. The analog-type operation with continuous current variation enables multi-level evaluation with more intermediate health-assessments such as ``low'', ``medium'', and ``high'' cautionary assessments. Combining the reconfigurable logic-type operation with the analog-type operation, dynamic health monitoring is possible. For instance, consider a scenario where either of the two diagnostic parameters fluctuates rapidly between ``high'' and ``normal'' within short time intervals. Such scenario is depicted in Figs.\hyperref[fig5]{5e,f} for heart-rate and blood-pressure, respectively, in which $I_{\text{out}}$ is recorded over time for alternate ``high'' and ``normal'' of the diagnostic parameter. Initially, $I_{\text{out}}$ remains low, indicating a non-critical state. However, after a series of ``high'' levels of the diagnostic parameter, $I_{\text{out}}$ crosses a defined threshold, triggering a medical emergency alert, an appropriate response to sustained abnormal conditions. In addition, the threshold-crossing point occurs at different number of ``high'' levels between the two diagnostic parameters. This time difference reflects the varying urgency associated with these parameters which could offer valuable insights for differential diagnosis. Next, the impact of dynamic input fluctuation is examined for a heart-patient (Fig.\hyperref[fig5]{5g}). Here, $I_{\text{out}}$ remains above the threshold current at all ``high'' and ``normal'' levels, maintaining a continuous medical emergency condition. These results demonstrate that our devices are capable of performing complex decision-making tasks, as exemplified through a physical health diagnostic model that integrates and responds to three physiological indicators. This highlights the potential of the system for real-time, adaptive health monitoring applications.

\section*{Outlook}\label{sec13}
We have demonstrated polymorphic functionalities of a LAO/STO heterostructures-based single nanowire via the manipulation of the q2-DEG at room temperature. These functionalities can be integrated to form complex circuits for applications in reservoir computing, neuromorphic systems, and reconfigurable synaptic logic. The tunability of the device enables on-demand modulation of device properties, offering versatile functionality within a single platform with advantage of its ease of fabrication and scalability. Our LAO/STO-based polymorphic oxide nanostructures thus pave the way for full oxide-based monolithic integrated circuits combining conventional sequential and non-conventional computing architectures. Its compatibility with Si-technology further supports the development hybrid CMOS-oxide architectures that incorporate memristive and memcapacitive functionalities with standard circuit elements. In particular, memcapacitor-based artificial neural networks enable ultra-low energy consumption in computing technologies \cite{demasius2021energy}. Alternatively, energy-efficient implementations such as spintronic logic devices can be realized using our LAO/STO system \cite{manipatruni2019scalable}, offering significantly improved scalability and logic density over conventional CMOS technology, leveraging its strong spin-orbit coupling. Our lateral side gate approach enables tunable control of the superconducting state \cite{reyren2007superconducting,li2011coexistence}, offering potential applications in Josephson junctions \cite{monteiro2017side}. Also, taking the advantage of lateral structure, one can add complexity in the device either electrically or optically for optoelectronic functionalities. Finally, the runtime programmable functionalities within a simple oxide-based device, combined with their versatile applications, establishes a promising pathway towards fully integrable energy efficient polymorphic and neuromorphic computing devices.

\bibliography{sn-bibliography}
\section*{Methods}
\subsection*{Device fabrication}
The devices were fabricated in main three steps. In the first step, a TiO$_2$ terminated (001)-oriented STO substrate was spin-coated with negative photoresist. Then the device layout was patterned on the surface of the substrate using electron beam lithography followed by resist development. In the next step, 11-nm SiO$_2$ was deposited by electron beam evaporation and then a liftoff process was carried out to create a well defined structure. At last, pulsed laser deposition (PLD) was employed to grow 6 u.c. of LAO on the surface by ablating a single crystalline LAO target at a frequency of 1 Hz using a KrF excimer laser ($\lambda$ =248 nm) at a substrate temperature of 780 $^{\circ}$C and an oxygen partial pressure of 1$\times$10$^{-3}$ mbar followed by annealing at 500 $^{\circ}$C for 1 h in 500 mbar of oxygen pressure. Finally, the device is ready consisting of a nanowire and two side-gates where the q2-DEG is formed, while other regions remain insulating due to the growth of amorphous LAO on the previously deposited SiO$_2$. The details of device fabrication process can be found in ref. \cite{miller2021room}. If not specified otherwise, the investigated wire is around 1 $\mu$m long and 100 nm wide. The wire-gate distance is around 400 nm. 

\subsection*{Device characterization}
At first, ultrasonic bonding with Al-wire was performed to directly contact the q2-DEG for all the electrical measurements. For current-voltage measurements, a Keithley source meter (Model: 213 Quad Voltage Source) was used and current was evaluated by adding a resistor of 10 k$\Omega$ or 100 k$\Omega$ in the circuits and measuring voltage drop through the resistor by a Keithley multimeter (2000 series). A lock-in amplifier (EG$\&$G Instruments, model: 7265) was employed to measure the real and imaginary parts of the current in the circuit by applying an ac signal using a Keithley arbitrary waveform generator (Model: 3390) in addition to the dc signal and used to evaluate the capacitance value. The data were extracted using labView program. All presented measurements were conducted at room temperature and in the dark. 

To investigate the FET-operation, the lateral gates are connected to an external gate voltage ($V_{\text{G}}$). A bias voltage is applied to the top contact (designated as drain) while the bottom contact (source) is connected to the common ground. To switch the device functionality from transistor- to memristive operation, the laterally defined side-gates are set on a floating potential as illustrated in the middle panel of Fig.\hyperref[fig1]{1a}. Capacitance measurements were performed between the channel and one of the lateral gates, while the other gate was kept floating, as shown in right panel of Fig.\hyperref[fig1]{1a}. 

For physical RC application utilizing a 1T1MC integrated circuit, input voltage pulse is applied to the gates of the transistor, while the output voltage is acquired from the memcapacitor, as illustrated in Fig.\hyperref[fig2]{2a}. To demonstrate the pattern recognition of monochrome digit images using the circuit, the gray and white pixels are assigned to input pulse amplitudes of 3 and -2 V, respectively, with a fixed pulse width of 150 ms. 

The schematic of the 1T1M structure for synaptic operation is shown in Fig.\hyperref[fig3]{3a} where a presynaptic voltage ($V_{\text{G}}$) is applied to one gate of the transistor, and the PSC is measured from the memristor. The other transistor gate is left floating to facilitate charge trapping and de-trapping, thereby modulating the conductivity to emulate the operation of a synaptic transistor \cite{qian2017multi}.

For the OR gate operation two transistors are connected in parallel with one memristor in series, while for the AND gate, two transistors are connected in series with the memristor, as illustrated in the insets of Figs.\hyperref[fig4]{4a,b}, respectively. A drain voltage ($V_{\text{D}}$) of 4 V was kept fixed during all the logic operations and their memory investigation. The reconfigurability of logic operations between AND and OR gate is realized considering another threshold current of 40 nA to distinguish between logic output ``0'' and ``1''.

\section*{Acknowledgements}

The authors gratefully acknowledge financial support from the state of Bavaria and the Deutsche Forschungsgemeinschaft (DFG, German Research Foundation) under Germany’s Excellence Strategy through the Würzburg-Dresden Cluster of Excellence on Complexity and Topology in Quantum Matter “ct.qmat” (EXC 2147, Project ID 390858490) as well as through the Collaborative Research Center SFB 1170 “ToCoTronics” (Project ID 258499086). VLR acknowledges the support from Conselho Nacional de Desenvolvimento Científico e Tecnológico (CNPq-Brasil) Proj. 311536/2022-0. 

\section*{Author contribution}
F.H. and S.H. initiated and guided the study. M.Spring, J.G. and B.L. grew the sample in discussion with M.Sing and R.C., S.K. and M.K. fabricated the devices. K.M. initiated the experiment, S.P. designed and conducted the experimental work in discussion with F.H.. S.P, F.H., V.L. and S.H. analyzed and interpreted the experimental results. S.P. and F.H. wrote the manuscript, with input from all coauthors.
\section*{Data availability}
The data that support the findings of this study are available from the
corresponding authors upon reasonable request.

\section*{Declarations}

\bmhead{Additional information}

\textbf{Supplementary information} The online version contains supplementary
material available at...
\begin{itemize}
\item \textbf{Conflict of interest/Competing interests} 
The authors declare no competing interests.
\item \textbf{Correspondence and requests for materials} should be addressed to Soumen Pradhan (soumen.pradhan@uni-wuerzburg.de) and Fabian Hartmann (fabian.hartmann@uni-wuerzburg.de).

\end{itemize}



\section*{Supplementary material}

# Supplementary Material:

## Oxide Interface-Based Polymorphic Electronic Devices for Neuromorphic Computing

Soumen Pradhan[1*], Kirill Miller[1], Fabian Hartmann[1*], Merit Spring[2], Judith Gabel[2], Berengar Leikart[2], Silke Kuhn[1], Martin Kamp[1], Victor Lopez-Richard[3], Michael Sing[2], Ralph Claessen[2], Sven Höfling[1]

[1*]Julius-Maximilians-Universität Würzburg, Physikalisches Institut and Würzburg-Dresden Cluster of Excellence ct.qmat, Lehrstuhl für Technische Physik, Am Hubland, Würzburg, 97074, Bavaria, Germany.

[2]Julius-Maximilians-Universität Würzburg, Physikalisches Institut and Würzburg-Dresden Cluster of Excellence ct.qmat, Experimentelle Physik 4, Am Hubland, Würzburg, 97074, Bavaria, Germany.

[3]Universidade Federal de São Carlos, Departamento de Fı́sica, Street, São Carlos, 13565-905 , SP, Brazil.

*Corresponding author(s). E-mail(s):

soumen.pradhan@uni-wuerzburg.de.

fabian.hartmann@uni-wuerzburg.de.

**Supplementary Note 1:**

The role of symmetry of bias voltage with respect to the middle of the nanowire device can be confirmed by means of a push-pull measurement. Here, the voltages $V_D$ and $V_S = -V_D$ are applied to the drain and source contacts, respectively. During the measurement, the voltage, $V_D$ is swept to $-V_S$ and $V_S$ from $-V_S$ to $V_D$. Figure S1 shows the corresponding current-voltage

characteristics. The push-pull method provokes an almost point symmetrical characteristic around zero bias voltage, and the memristive response is of a type-II character, as one expects in the case of symmetric carrier trapping processes[1]. For the push-pull measurement with grounded gates, the maximum and minimum currents are 2.6 µA and -2.3 µA, respectively. The small current asymmetry is most likely caused by a slightly asymmetrical device layout, unavoidably induced by the different processing steps. The current-voltage characteristic for grounded gates indicates a small potential barrier in the channel (inset: Figure S1). For the floating gate operation, we observe a high resistance plateau around zero bias voltage (inset: Figure S1). For the down-sweep (i.e., from +5 V to – 5 V), the characteristic shows symmetrical threshold voltages with $\Delta V_{th}^{down}$ = 0.78 V. For the up-sweep, the measured voltage difference increases up to $\Delta V_{th}^{up}$ = 1.22 V. The type-II memristive response is discernible by the area enclosed during a bias voltage sweep. Therefore, the memory content in the device is highly sensitive on not only the amplitude of $V_D$ but also the symmetry of bias between the drain and source contact.

**Supplementary Note 2:**

To understand the output inefficiency for higher $V_D$ in the one transistor one memcapacitor (1T1MC) circuit, $V_D$ is swept from 0 to 4 V keeping the transistor switched 'on' at a fixed gate voltage of 3 V, as represented in Fig. S3. The results reveal that the output voltage ($V_O$) increases linearly with $V_D$ upto ≈2.5 V, beyond which the slope changes, and the reverse sweep exhibits hysteresis. It can be explained as a tuning of the local voltage efficiency. Since $V_G$=3 V, a polarity inversion is expected in the potential difference at $V_D$≈3 V. We note that all these capacitances change drastically with polarity.

**Supplementary Note 3:**

As shown in Fig.2b in the main text, $V_O$ depends on the input pulse width in a 1T1MC circuit. Therefore, to demonstrate the reservoir computing (RC) application of the circuit configuration, the width of the input pulses also plays a significant role. In the main text, the width of the pulses were kept fixed at 150 ms for the monochrome digit recognition task and all the 16 possible 4-bits pulse trains. Here, we measured $V_O$ for all 16 states for pulse widths of 250 ms, 350 ms, 500 ms and 1 s, as shown in Fig. S5. The variation in reservoir states with different pulse widths reflects dynamic learning behavior, opening path for time series analysis.

**Supplementary Note 4:**

In human brain, continued or repeated stimulation results in long term storage of information. To mimic the same behavior, a presynaptic signal is applied with varying the width and number of the voltage pulses to the one transistor one memristor (1T1M) device structure as shown in Fig.S6a. Figure S6b shows the input voltage ($V_G$) pulse of 3 V of different width and corresponding post synaptic current (PSC) for fixed $V_D$ of 4 V. It is observed that, there is an increase in PSC at the end of each pulse which corresponds to the increment of synaptic strength. Moreover, after returning the voltage pulse to 0 V, there is only a slow variation of PSC towards a relaxation value. Now, to better understand the results, change in PSC (ΔPSC) between beginning and end of a pulse is extracted and shown for different pulse width. With increase in pulse width, ΔPSC increases in a non-linear way which indicates the transition from short term memory (STM) to long term memory (LTM) in the device. Next, variation of input pulse number and corresponding PSC are shown in Fig. S6d. Similarly, PSC enhances with increase in pulse number. Here, the synaptic strength is extracted by evaluating the ratio of PSC of $N^{th}$ pulse and $1^{st}$ pulse ($PSC_N/PSC_1$) for each pulse train. Non-linear variation of $PSC_N/PSC_1$ with pulse number shown in Fig. S6e confirms the tuning possibility of synaptic strength like human brain. Also, before applying every pulse, PSC is reset to zero by applying a reset voltage of -2 V which is similar to resetting the memory of brain.

**Supplementary Note 5:**

The operation of a NOT gate is demonstrated using the device in transistor mode and a very high resistor (100 MΩ) compared to the "on" state resistance of the transistor in series as shown in Fig. S7a. The gate voltage ($V_G$) of the transistor is used as input signal and voltage between the resistor and transistor is considered as output signal ($V_O$). The drain voltage ($V_D$) of the transistor is kept fixed at 1 V during the logic operation. Now, to demonstrate the logic operation, $V_G$ is swept from -1 V to 1 V. As depicted in Fig. S7b, when $V_G$ is negative, $V_O$ is high with a value ~0.95 V at $V_G$ = -1 V, then $V_O$ decreases drastically when $V_G$ approaches 0 V and at $V_G$ = 1 V, $V_O$ reaches ~0.05 V. Now, considering $V_G$ of -1 V and 1 V as logic input "0" and "1", respectively, and $V_O$ = 0.5 V as threshold voltage to distinguish between logic output

"0" and "1", the operation of a NOT gate can be confirmed in a one transistor one resistor (1T1R) circuit.

**Supplementary Note 6:**

The logic computations and in-situ memory of logic output are demonstrated in two transistors one memristor (2T1M) device configuration utilizing one of the lateral gates from the two transistors as two input signals (Fig.S8a). To understand the origin of memory, gate voltages are swept between minimum and maximum values in forward and reverse directions [2]. Figures S8b,c show the output current ($I_{out}$) for a gate voltage ($V_G$) sweep from -2 to 3 V and then 3 to -2 V of one transistor while $V_G$ of the other transistor is kept fixed at -2 V for constant $V_D$ of 4 V. It is observed that there is an anti-clockwise hysteresis in $I_{out}$. This is because the channel opens at around -1 V during forward sweep direction, and slowly current increases as we increase $V_G$, and at $V_G \sim 0$ V, we see some current flow. In contrast, during reverse sweep cycle, a higher current is observed through the channel at $V_G = 0$ V which is due to the discharging of the floating gates in the whole circuit. Therefore, hysteresis in transfer characteristics is expected. Now, to confirm this, we have changed the $V_G$ sweep rate and with increasing sweep rate there is a reduction in hysteresis since less time is spent above threshold voltage for faster sweep cycle to discharge charges to the channel from the floating gates.

**Supplementary Note 7:**

We conducted $I_{out}$-$V_D$ full-cycle measurements between ±4 V using the circuit configuration shown in Fig. S9a, with input combinations-00, 10, 01, and 11. Here, the input "0" corresponds to $V_G$=-2 V and input "1" refers to $V_G$=3 V. As expected, hysteresis is observed in all cases except for the positive $V_D$ sweep corresponding to IN-00, where the output current remains very low (see Fig.S9). Notably, during up-sweep at $V_D$ =3 V, the $I_{out}$ for IN-11 is significantly higher compared to IN-10 and 01, while during down sweep, all three inputs exhibit high $I_{out}$ values. Therefore, by selecting an appropriate threshold current between these two sets of output levels, it is possible to construct an AND gate from an existing OR gate within the same circuit configuration. However, the AND gate operation is also shown from $I_{out}$-$V_D$ full-cycle measurements between ±4 V using the circuit configuration shown in Fig.S10a. As we observe, there is negligible current flow in the positive side of $V_D$ sweep cycle for IN-00, 10 and 01 (Fig.S10b). This is because the channel is depleted for these input combinations, resulting in "low" logic output. On the other hand, only for IN-11, the channel becomes open and a relatively high current is observed, making a "high" logic output.

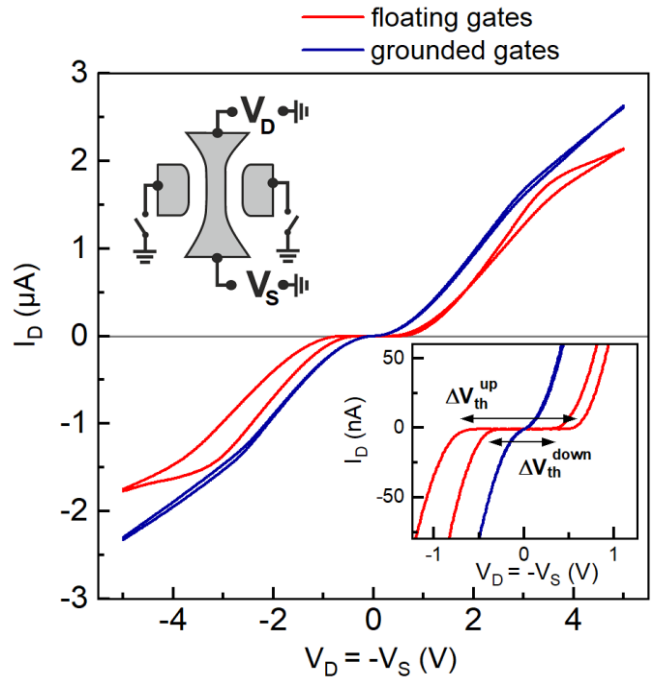

Figure S1: Current-voltage characteristics for the push-pull measurement with both the gates at floating and grounded conditions. The voltage $V_D$ is applied to the drain contact and $V_S$ (= - $V_D$) to the source contact. The current-voltage characteristic can be symmetrized and the memristor response can be tuned from type-I to type-II.

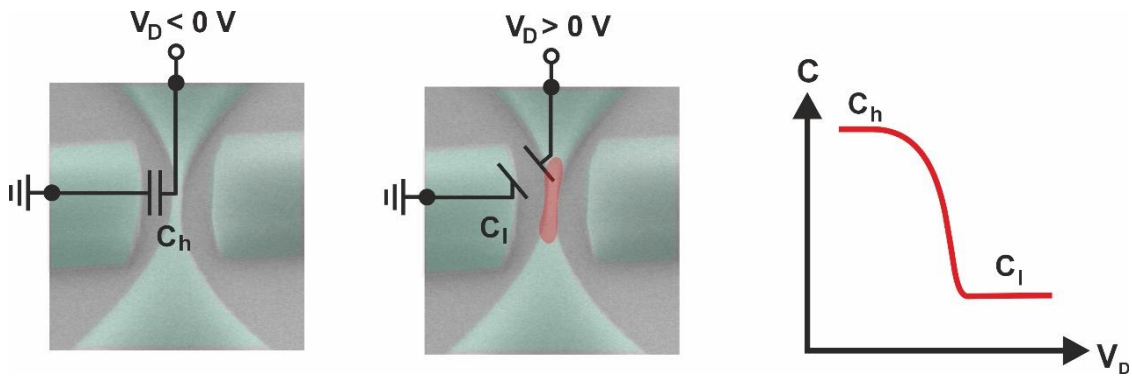

Figure S2: Capacitance representation under reverse and forward bias ($V_D$) condition in the memcapacitor configuration with schematic representation of transition of capacitance between high ($C_h$) and low ($C_l$) value with varying $V_D$ between negative and positive voltages.

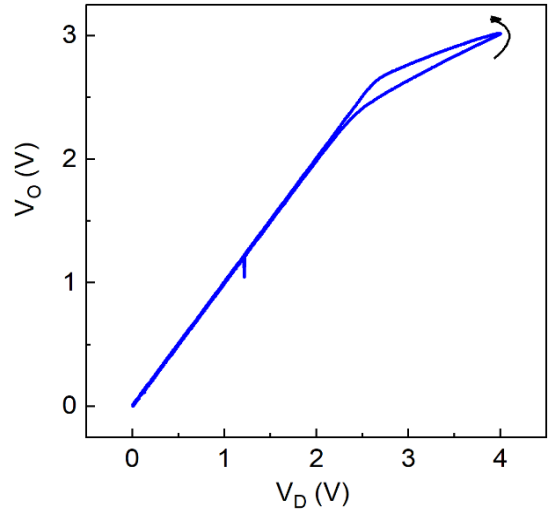

Figure S3: Hysteresis in $V_O$ with $V_D$ triangular sweep cycle between 0 and 4 V keeping the transistor 'on' with $V_G$ of 3 V in 1T1MC circuit configuration.

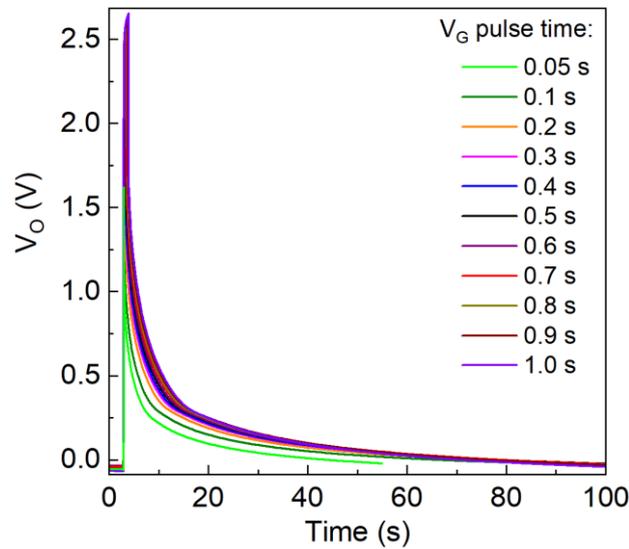

Figure S4: Output voltage ($V_O$) by applying single input pulse between -2 and 3 V of varying pulse width from 50 ms to 1 s at fixed $V_D$ of 4 V for 1T1MC circuit configuration. The decrement of $V_O$ towards 0 V due to leakage of charges in the memcapacitor after switching off the transistor takes place in two steps: initially a very fast fall down followed by a comparatively slow process.

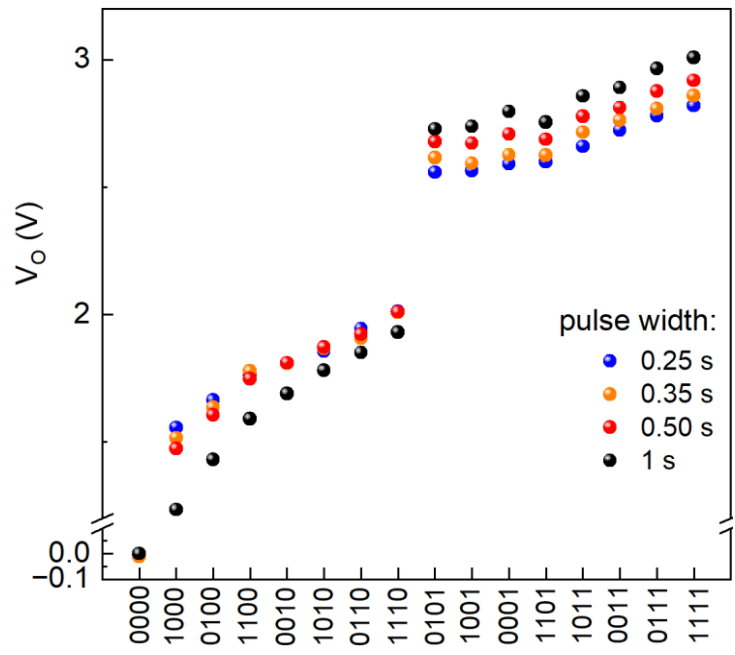

Figure S5: $V_O$ at the end of all 16-types of 4-bit pulse trains applied to the input in 1T1MC circuit for different pulse width at fixed $V_D$ of 4 V. Different pulse width corresponds to a different reservoir state, with almost all 16-states being distinguishable.

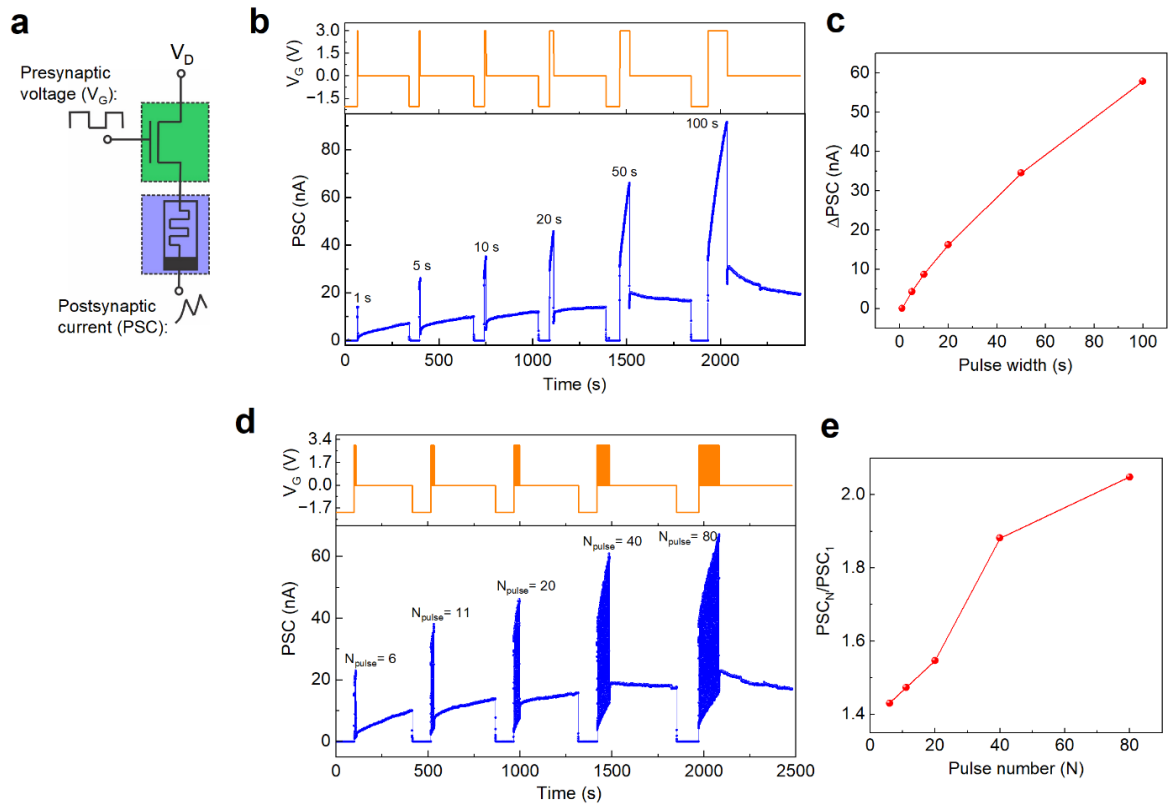

Figure S6: **Transition from STP to LTP in 1T1M device: a,** schematic diagram of 1T1M device configuration with presynaptic voltage pulse to the gate of transistor and postsynaptic current (PSC) from the memristor, conversion from STP to LTP by increasing the input pulse **b,** width and **d,** number between -2 and 3 V at constant $V_D$ of 4 V, **c,** plot of ΔPSC (difference between final and initial current during stimulation of a pulse) with pulse width and **e,** plot of $PSC_N/PSC_1$ with pulse number.

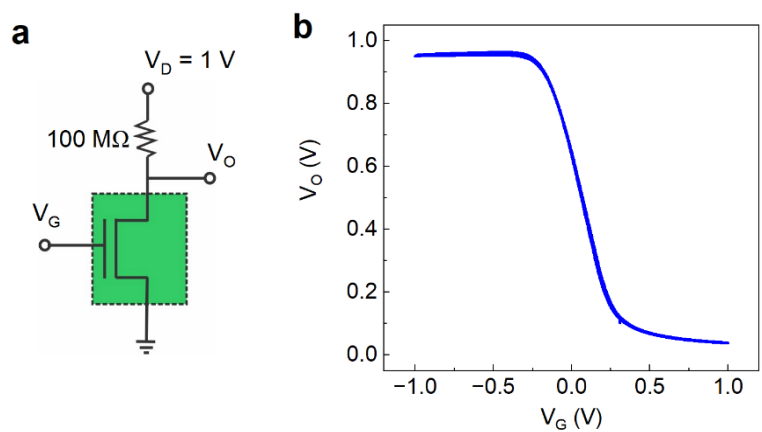

Figure S7: **NOT gate operation in 1T1R device: a,** schematic illustration of 1T1R device configuration where one transistor is connected in series with one comparatively very high resistor; **b,** variation of output voltage ($V_O$) collected between the resistor and transistor as a function of gate voltage ($V_G$) at fixed $V_D$ of 1 V. The data represents operation of a NOT gate.

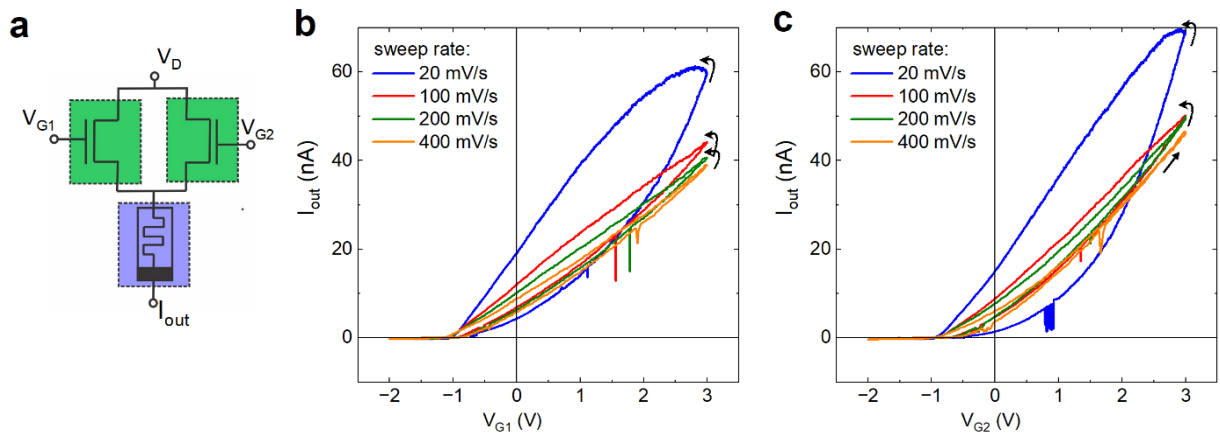

Figure S8: **Memory in transfer characteristics in 2T1M device: a,** schematic diagram of 2T1M device configuration where two transistors, connected in parallel, are joined with one memristor in series, hysteresis in transfer characteristics curves for **b,** 1st transistor (left) keeping 2nd transistor (right) switched 'off' at $V_{G2}$= -2 V and **c,** 2nd transistor keeping 1st transistor switched 'off' at $V_{G1}$= -2 V for different sweep rate at fixed $V_D$ of 4 V.

9

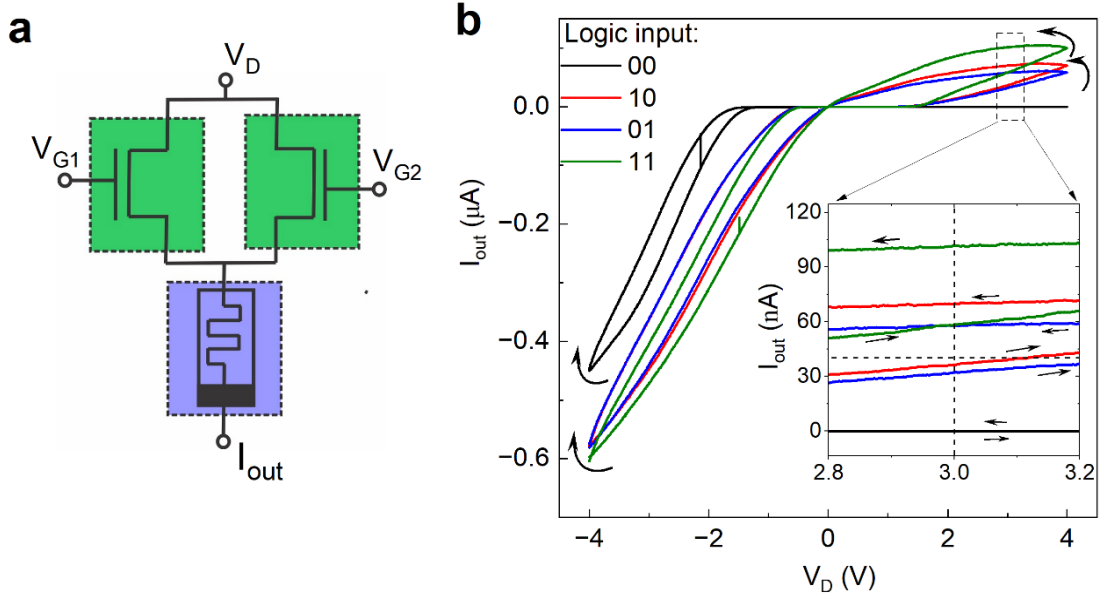

Figure S9: **Logic OR operation in 2T1M device: a,** schematic diagram of 2T1M circuit diagram where two transistors in parallel configuration are connected with one memristor in series, and corresponding **b,** output current ($I_{out}$) with $V_D$ sweep between ± 4 V for logic input "00" ($V_{G1}= V_{G2}=-2$ V), "10" ($V_{G1}=3$ V , $V_{G2}=-2$ V), "01" ($V_{G1}=-2$ V , $V_{G2}=3$ V) and "11" ($V_{G1}= V_{G2}=3$ V). In the positive side of $V_D$ there is increment of current for logic inputs "10", "01" and "11" but not for "00" which confirms the logic OR operation. Inset shows the zoom-in plot around $V_D = 3$ V where forward sweep cycle for "10" and "01" can be distinguished from reverse sweep cycle for "10", "01" and "11" assuming a threshold current of 40 nA to be used for reconfigurable logic operation.

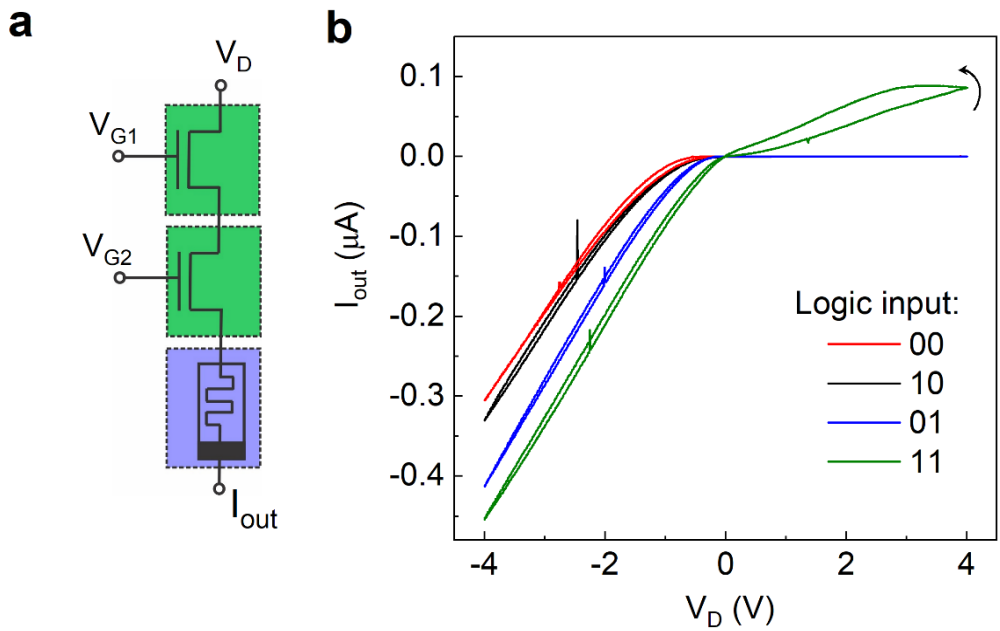

Figure S10: **Logic AND operation in 2T1M device: a,** schematic diagram of 2T1M circuit diagram where two transistors and one memristor are connected in series, and corresponding **b,** output current ($I_{out}$) with $V_d$ sweep between ± 4 V for logic input 00 ($V_{G1}$= $V_{G2}$=-2 V), 10 ($V_{G1}$=3 V , $V_{G2}$=-2 V), 01 ($V_{G1}$=-2 V , $V_{G2}$=3 V) and 11 ($V_{G1}$= $V_{G2}$=3 V). In the positive side of $V_D$ there is increment of current only for logic input 11 which confirms the logic AND operation.


\end{document}